\documentclass[12pt,preprint]{aastex}

\shorttitle{Massive dust shell around SN~2002hh}
\shortauthors{Barlow et al.}

\begin{document}


\title{Detection of a massive dust shell around the Type~II supernova
SN~2002hh}


\author{M.~J.~Barlow\altaffilmark{1}, B.~E.~K.~Sugerman\altaffilmark{2}, 
J.~Fabbri\altaffilmark{1},
M.~Meixner\altaffilmark{2}, R.~S.~Fisher\altaffilmark{3}, 
J.~E.~Bowey\altaffilmark{1}, N.~Panagia\altaffilmark{2},
B.~Ercolano\altaffilmark{1}, G.~C.~Clayton\altaffilmark{4}, 
M.~Cohen\altaffilmark{5},
T.~M.~Gledhill\altaffilmark{6}, K.~Gordon\altaffilmark{7}, 
A.~G.~G.~M.~Tielens\altaffilmark{8},
A.~A.~Zijlstra\altaffilmark{9}}


\altaffiltext{1}{Dept. of Physics \& Astronomy, University College London, 
Gower Street, London WC1E~6BT, UK}
\altaffiltext{2}{Space Telescope Science Institute, 3700 San Martin Drive
Baltimore, MD 21218}
\altaffiltext{3}{Gemini Observatory, Northern Operations Center, 670 North
A'ohoku Place, Hilo, HI 96720}
\altaffiltext{4}{Department of Physics and Astronomy, Louisiana State
University, Baton Rouge, LA 70803}
\altaffiltext{5}{Radio Astronomy Laboratory, University of California at
Berkeley, Berkeley, CA 94720}
\altaffiltext{6}{Department of Physics, Astronomy \& Maths, University of
Hertfordshire, College Lane, Hatfield AL10~9AB, UK}
\altaffiltext{7}{Steward Observatory, University of Arizona, Tucson, 
Arizona 85721}
\altaffiltext{8}{Kapteyn Astronomical Institute, P.O. Box 800, 9700 AV
Groningen, The Netherlands}
\altaffiltext{9}{School of Physics \& Astronomy, University of
Manchester, PO Box~88, Manchester M60~1QD}


\begin{abstract} Dust emission from the Type~II supernova SN~2002hh in 
NGC~6946 has been
detected at mid-infrared wavelengths by the {\em Spitzer Space Telescope}
from 590 to 758 days after outburst and confirmed by higher angular
resolution Gemini-N mid-IR observations. The day-600 $5.8-24$-$\mu$m
emission can be fit by a 290-K blackbody having a luminosity of
1.6$\times10^7$~L$_{\odot}$. The minimum emitting radius of 
1.1$\times10^{17}$~cm is
too large for the emitting dust to have been formed in the supernova
ejecta. Using radiative transfer models and realistic dust grain
parameters, fits to the observed flux distribution could be obtained with
an optically thick dust shell having a mass of 0.10-0.15~M$_\odot$,
corresponding to a total dust+gas mass in excess of 10~M$_\odot$,
suggesting a massive M~supergiant or luminous blue variable precursor
to this self-obscured object.

\end{abstract}




\section{Introduction}

\noindent
During their giant phase of evolution, low and intermediate mass stars are
known to be important contributors to the interstellar dust budget.
However, massive stars, in particular their supernovae (SNe), could make
the dominant contribution to the dust budget of galaxies (Clayton 1979;
Dwek \& Scalo 1980; Todini \& Ferrara 2001). Evidence for dust formation
by at least some SNe comes from precise isotopic abundance ratio studies
of grain inclusions found in meteorites (e.g. Clayton, Amari \& Zinner 
1997; Travaglio et al. 1999). A determination of the dust contribution from
core-collapse SNe (Types II, Ib and Ic) could have widespread consequences,
in particular for our understanding of the origin and evolution of
galaxies in the early universe. Many of the earliest-formed galaxies are
extremely dusty, as evidenced by the detection of their highly-redshifted
infrared (IR) emission at submillimeter wavelengths (e.g. Ivison et al.
2000). Massive stars in the starbursts that power these young galaxies are
the most plausible sources of this dust.

Model studies have shown that the mid-IR spectral region
(5-30~$\mu$m) is particularly suitable to trace the onset of dust
formation in SN ejecta and to determine the amount of dust formed (e.g.
Dwek 1988; Kozasa, Hasegawa \& Nomoto 1989; Todini \& Ferrara 2001;
Clayton, Deneault \& Meyer 2001). These dust formation models predict that
emission from ejecta condensates should become detectable at mid-IR
wavelengths within 1-2 years of outburst. SN~1987A, in the Large
Magellanic Cloud, was estimated to have formed a modest amount of dust in
its ejecta (Dwek et al. 1992; Wooden et al. 1993), with the dust believed
to have first condensed about 530 days after outburst (Danziger et al.
1991). In contrast, the level of submm emission from the 300-year old
SN remnant Cas~A has been interpreted as indicating the presence of at
least two solar masses of cold dust within the remnant (Dunne et al. 2003).
However, amongst the grounds given by Dwek (2004) for disputing this dust 
mass estimate was that it would exceed the mass of refractory 
elements in the ejecta of the likely progenitor star, while
Krause et al. (2004) argued that the submm emission
detected towards Cas~A could originate from a foreground molecular cloud
complex, leaving still uncertain
the observational case for SNe as major dust contributors
to galaxies. Here we report the detection by our SEEDS collaboration 
(Survey for Evolution of Emission from Dust in Supernovae) of 
thermal dust
emission from the Type~II SN~2002hh some 600 days after its discovery,
based on mid-infrared observations made by the {\em Spitzer Space
Telescope (SST)}, together with follow-up mid-IR observations obtained
with the 8-m Gemini-North telescope.

\section{Observations}

The Type~II supernova SN~2002hh is located in the spiral galaxy NGC~6946,
at a distance of 5.9$\pm$0.4~Mpc (Karachentsev, Sharina \& Huchtmeier 
2000). NGC~6946
was observed by the SINGS Legacy program (Kennicutt et al. 2003) with the
{\em SST}'s InfraRed Array Camera (IRAC) at 3.6, 4.5, 5.8 and 8.0~$\mu$m,
on June 10th 2004 and again on November 25th 2004, 590 days and 758 days,
respectively, after the October 27th 2002 discovery of SN~2002hh (Li
2002). Both IRAC observation sequences took 35 minutes, yielding
the effective exposure times per pixel that are listed in Table~1. The 
galaxy was also observed at 24, 70 and 160~$\mu$m, for 55.2 minutes on 
both July 9th and 11th 2004, with the 
Multiband Imaging Photometer for Spitzer (MIPS). 
Pipeline-calibrated images of NGC~6946 were obtained from the {\em SST}
public archive. Figs.~1(a) and (b) show IRAC 5.8-$\mu$m and 8.0-$\mu$m
images of a 30$''$ by 29$''$ region around SN~2002hh, obtained with a
pixel size of 1$''.$1, from the June 10th 2004 SINGS observation of
NGC~6946. They clearly show the SN (marked as Star 1) and a bright
adjacent field star (Star 2). Using crowded-field PSF subtraction, we
removed these two sources, revealing three additional sources, which we
named Stars 3, 4 and 5. These sources were identified in all four IRAC
bands, with the fluxes listed in Table~1. The error estimates listed
there include allowances for flat-field errors, profile errors in fitting 
the PSF, readnoise and poisson noise. Astrometry for Stars 1-5 from
these IRAC images was reported by Barlow et al. (2004). The J2000
coordinates measured for SN~2002hh are R.A. = 20$^{\rm h}34^{\rm m}44^{\rm
s}$.23, Decl. = +60$^{\rm o}07'19''.5$ ($\pm0''.3$ in each coordinate).
These agree well with the J2000 radio position of SN~2002hh measured
by Stockdale et al. (2002), R.A. = 20$^{\rm h}34^{\rm m}44^{\rm
s}$.25, Decl. = +60$^{\rm o}07'19''.4$ ($\pm0''.2$ in each coordinate).

From its 2MASS and IRAC colours, Star 2 (= 2MASS~20344320+6007234) is
likely to be a foreground cool giant in the Milky Way. Stars 4 and 5 were
suspected to be a single extended source, possibly an H~{\sc ii} region;
for a distance of 6~Mpc to NGC~6946, Star 4 has a projected separation of 
90~pc from SN~2002hh. Fig.~1(c) shows the MIPS 24-$\mu$m image of the same
region, obtained with a pixel size of $2''.6$. The supernova (Star 1) and
Stars 4-5 are clearly detected at 24~$\mu$m, with the fluxes listed in
Table~1. The IRAC and MIPS fluxes listed in Table~1 have not been
corrected to take account of intrinsic source colours; such corrections
are estimated to be small. The listed flux uncertainties correspond to the
photometric statistical errors; the absolute calibration uncertainties for
IRAC and MIPS are estimated to be 5-6\%. It is noteworthy that the
8-$\mu$m flux from from the supernova (Star 1) showed a 25\% drop between
the IRAC observations of June and November 2004.

The supernova is only marginally resolved from its neighbouring sources in
the IRAC and MIPS images, due to the relatively low angular resolution of
the 85-cm {\em SST} ($2''.4$ at 8~$\mu$m). To address the issue of source
confusion, we used the Michelle mid-IR imager/spectrometer on the 8-m
Gemini-North telescope to obtain an 11.2-$\mu$m image of 2002hh on
September 26th 2004, with an on-source observation time of 900~s.  
Fig.~1(d) shows the region around SN~2002hh in
the Michelle image. With $\sim10$ times the spatial resolution of the IRAC
observations, the Michelle image completely resolves SN 2002hh from its
neighbors and confirms that mid-IR emission originates from the location
of the supernova. Both the SN and Star 2 are easily detected in this
image, while Star 4/5 is confirmed to be an extended source, which peaks
in brightness to the south-east.

\section{The IR spectral energy distribution and the mass of emitting 
dust}

The observed June/July 2004 MIPS/IRAC spectral energy distribution of
SN~2002hh is shown in Fig.~2. The longer wavelength {\em SST} data can be
fitted by a 290~K blackbody, which for a distance of 6~Mpc corresponds to
a luminosity of 1.6$\times10^7$~L$_{\odot}$ and an emitting radius of
1.1$\times10^{17}$~cm. More realistic dust particles have
$\lambda^{-\alpha}$ emissivities in the IR, with $\alpha$ typically
between 1--2. For an $\alpha$ = 1 emissivity law, a fit to the June/July
2004 8- and 24-$\mu$m fluxes gives a grain temperature of 225~K and a
characteristic emitting radius of 5$\times10^{17}$~cm, too large for the
emitting dust to have been formed in the main SN ejecta, since material
traveling at a bulk velocity of 6000~km~s$^{-1}$ would take 26 years to
reach this radius.  Light from the supernova would take 6 months to reach
this radius; the light crossing time of one year for a shell of the
same radius appears consistent with the observed flux decrease at 8-$\mu$m
of 25\% in 5.5~months (Table~1). We infer that the observed IR emission
from SN~2002hh is from pre-existing circumstellar (CS) dust ejected by the
supernova progenitor star. From day-20 near-IR spectra, an extinction of
A$_{\rm V}$ = 6.1 mags has been measured towards the SN by Meikle et al.
(2002) of which $\sim$1.1 mags were estimated by them to be due to
foreground dust in the Milky Way. At SN~2002hh's angular separation of
129$''$ from the nucleus, an internal extinction of A$_{\rm V} =
1.2{+1.3\atop -1.7}$ magnitudes to the midplane of NGC~6946 is predicted
(Holwerda et al. 2005). Thus a large fraction of the 5 magnitudes of
extinction within NGC~6946 towards SN~2002hh appears to be due to CS dust,
which may have condensed within a stellar wind from an earlier
M~supergiant or Luminous Blue Variable (LBV) phase of evolution of the
progenitor star.

We have constructed a number of dust shell models that match the observed
June/July 2004 IR flux distribution from SN~2002hh. Fits were obtained for
total dust masses of 0.10-0.15~M$_\odot$, with visual optical depths,
$\tau_{\rm V}$, in the range 3--4, illuminated by central sources with
luminosities of (2.0--2.3)$\times10^7$~L$_\odot$ and effective
temperatures in the range 5000--7500~K.  The dashed line in Fig.~2
corresponds to an amorphous carbon model for a dust shell with an r$^{-2}$
density distribution, an inner shell radius $R_{in}$ =
4.0$\times10^{17}$~cm, a ratio, Y, of outer to inner shell radii of 3.0, a
radial optical depth $\tau_{\rm V}$ (= A$_{\rm V}/1.086$) of 4.0 and a
total dust mass of 0.10~M$_\odot$. The grains had a Mathis, Rumpl \&
Nordsieck (1977; MRN) $a^{-3.5}$ size distribution, with minimum and
maximum grain radii of $a_{min} = 0.005~\mu$m and $a_{max} = 0.25~\mu$m.
The model was calculated using dusty-MOCASSIN (Ercolano, Barlow \& Storey
2005), a 3-D Monte Carlo radiative transfer code, with the amorphous
carbon optical constants taken from Hanner (1988). The central source used
for the model had a luminosity of $L = 2.3\times10^7$~L$_\odot$ and an
effective temperature $T_{\rm eff}$ = 7500~K.The dotted line in Fig.~2
shows an r$^{-2}$ density distribution dust shell model calculated using
the 2-D radiative transfer code 2-Dust (Ueta \& Meixner 2003) for a dust
shell containing 25\% amorphous carbon and 75\% amorphous silicate grains
by mass, with MRN $a^{-3.5}$ size distributions and $a_{min} = 0.005~\mu$m
and $a_{max} = 1.0~\mu$m. The amorphous carbon and amorphous silicate
optical constants were taken from Zubko et al. (1996) and Draine \& Lee
(1984), respectively. The dust shell had $\tau_{\rm V}$ = 3, a mass of
0.15~M$_\odot$, R$_{in}$ = 2.1$\times10^{17}$~cm and Y = 1.5 and was
powered by a central source with $L = 2.1\times10^7$~L$_\odot$ and $T_{\rm
eff}$ = 7000~K. The best-fitting model (dash-dotted line in Fig.~2; also
calculated with 2-Dust) was an amorphous carbon and silicate model having
the same central source luminosity and optical depth as above, but with a
dust mass of 0.10~M$_\odot$, $R_{in}$ = 1.0$\times10^{17}$~cm, Y = 100 and
an embedded constant density shell or ``superwind'' located between
$R_{in}$ and 10$R_{in}$, followed by a smooth transition to an r$^{-2}$
density distribution having 1/25th of the density of the constant density
shell at the 10$R_{in}$ transition point.

For the best-fitting model (dash-dot line in Fig.~2), the adopted 25:75
ratio by mass for the carbon and silicate dust components is similar to
that found necessary to match the light echoes observed from the CS dust
closest to SN~1987A (Sugerman et al. 2005). Spectral features attributable
to both O-rich and C-rich dust species have been found in the IR spectra
of the CS dust shells around the candidate Galactic LBVs AFGL~2298 (Ueta
et al. 2001) and HD~168625 (O'Hara et al. 2003). Other grain species that
are relatively featureless in the mid-IR could be an alternative to carbon
grains, e.g. iron or iron oxide particles, or else `dirty' silicate grains
(Jones \& Merrill 1976), having larger optical and IR absorptivities
outside the mid-IR features than used here, could provide a match to the
observed flux distribution without the need for an additional grain
species.

\section{Discussion}

The CS dust mass of 0.10--0.15~M$_\odot$ around SN~2002hh is large,
corresponding to an associated gas mass of 10--15~M$_\odot$ if we assume a
typical interstellar gas to dust mass ratio of 100. From a fit to its ISO
IR spectrum, 0.17~M$_\odot$ of dust has been estimated to be around the
archetypal Luminous Blue Variable (LBV) $\eta$~Car (Morris et al. 1999).
Another Galactic LBV, AG~Car, has a CS dust shell extending from 0.37 to
0.81~pc radius, with a dust mass of 0.25~M$_\odot$ and a gas to dust mass
ratio of only 32 (Voors et al. 2000). A mass of 0.1~M$_\odot$ has been
derived for the dust shell around the galactic LBV-candidate AFGL~2298
(Ueta et al. 2001), while evidence has been found for at least two past
superwind events in the dust shell around the self-obscured M~supergiant
NML~Cyg, with total dust masses of 0.02 and $>$0.3~M$_{\odot}$,
respectively (Bloecker et al. 2001). It seems plausible that the dust
shell observed around SN~2002hh may have originated from an episode of
enhanced mass loss by the massive star progenitor of this Type~II
supernova.

The IR luminosity of 2$\times10^7$~L$_\odot$ derived for SN~2002hh at
$\sim$day-600 is much higher than the likely luminosity of the supernova
ejecta at that epoch. However, given that the light travel timescale
across the dust shell is in excess of a year for the models plotted in
Fig.~2, this could be explainable in terms of the time lags associated
with the emission from flash-heated circumstellar dust. A number of
Type~II SNe have previously been found to exhibit near-IR
($\lambda<5\mu$m) excesses that could be attributed to emission from hot
dust, often with post-300d luminosities comparable to or larger than that
observed from SN~2002hh (see e.g. Gerardy et al. 2002). The high
luminosities and the pattern of time evolution of these near-IR excesses
has led to them usually being interpreted as arising from flash-heated
pre-existing circumstellar dust shells (Dwek 1983; Graham \& Meikle 1986;
Gerady et al. 2002; Pozzo et al. 2004). The 2.2--3.5-$\mu$m emission
observed from SN~1979C between days 259 and 440, and from SN~1980K between
days 215 and 357, was interpreted by Dwek (1983) as due to pre-existing
dust, extending in both cases from $(3-8)\times10^{17}$~cm, with visual
optical depths of 0.3 and 0.03 and total dust masses of
$(0.6-3)\times10^{-2}$~M$_{\odot}$ and $(0.7-3)\times10^{-3}$~M$_{\odot}$,
respectively. Pozzo et al. (2004) estimated a dust mass of at least
$2\times10^{-3}$~M$_{\odot}$ around the Type~IIn SN~1998S, though they
suggested that both pre-existing and newly formed dust grains may have
played a role in the evolution of its near-IR spectral energy
distribution.

Our observations do not rule out the possibility that significant
quantities of new dust may have formed in the ejecta of SN~2002hh itself,
merely that the observed IR emission is dominated by dust located much
further out than the SN ejecta. At the distance of NGC~6946, the day-615
8-$\mu$m emission from the ejecta dust of SN~1987A (Wooden et al. 1993;
Bouchet \& Danziger 1993) would have been 35 times weaker than that
observed from SN~2002hh. SN~1987A's dust visual optical depth at day-600
was estimated to be 0.6 (Lucy et al. 1991), so even if it had been able to
condense much more dust than the few$\times10^{-4}$~M$_\odot$ estimated
(Dwek et al. 1992; Wooden et al. 1993), its mid-IR emission could only
have increased by a further factor of two, still leaving it a factor of
ten lower than observed from the CS shell surrounding SN~2002hh. One
indication of new dust forming in SN ejecta, generally seen in the second
year after outburst, is the development of asymmetric blue-shifted
emission-line profiles, caused by dust preferentially extinguishing
redshifted emission from material behind the supernova (Danziger et al.
1991). SN~2002hh has not so far shown any such line-profile evolution. A
comparison of visible spectra taken in July 2003 (Mattila, Meikle \&
Greimel 2004) with those taken in August and September 2004 (Clayton \&
Welch 2004) indicates that the H$\alpha$ profile shape did not undergo any
significant change between $\sim$250d and 700d after outburst.

Our IR observations of SN~2002hh indicate that significant amounts of dust
can be produced around the immediate progenitors of some massive-star
supernovae. The question of whether such dust can survive the impact of
high-velocity ejecta from the subsequent supernova and thereby go on to
enrich the dust content of the host galaxy is an interesting one.
SN~2002hh is located in a relatively nearby face-on galaxy, which enabled
it to be discovered in the optical despite its high dust obscuration. For
many galaxies, however, SNe with similar or even greater self-obscuration
could easily escape detection by optical supernova searches, suggesting
that IR-based supernova searches may be required in order to determine the
ratio of dusty to non-dusty SNe and the contribution made by the former to
the overall dust enrichment rate of galaxies. \\

\acknowledgments

We thank the Gemini Observatory for the 
award of Director's Discretionary Time for observations of SN~2002hh.
B. Sugerman has been supported by a Spitzer GO-P3333 grant and by
internal STScI-DDRF funding. M. Meixner has been supported by
NASA grant NAG 5-11460. J. Bowey and J. Fabbri were supported
by grants from the United Kingdom Particle Physics and Astronomy 
Research Council.

\clearpage

\begin{figure}
\plotone{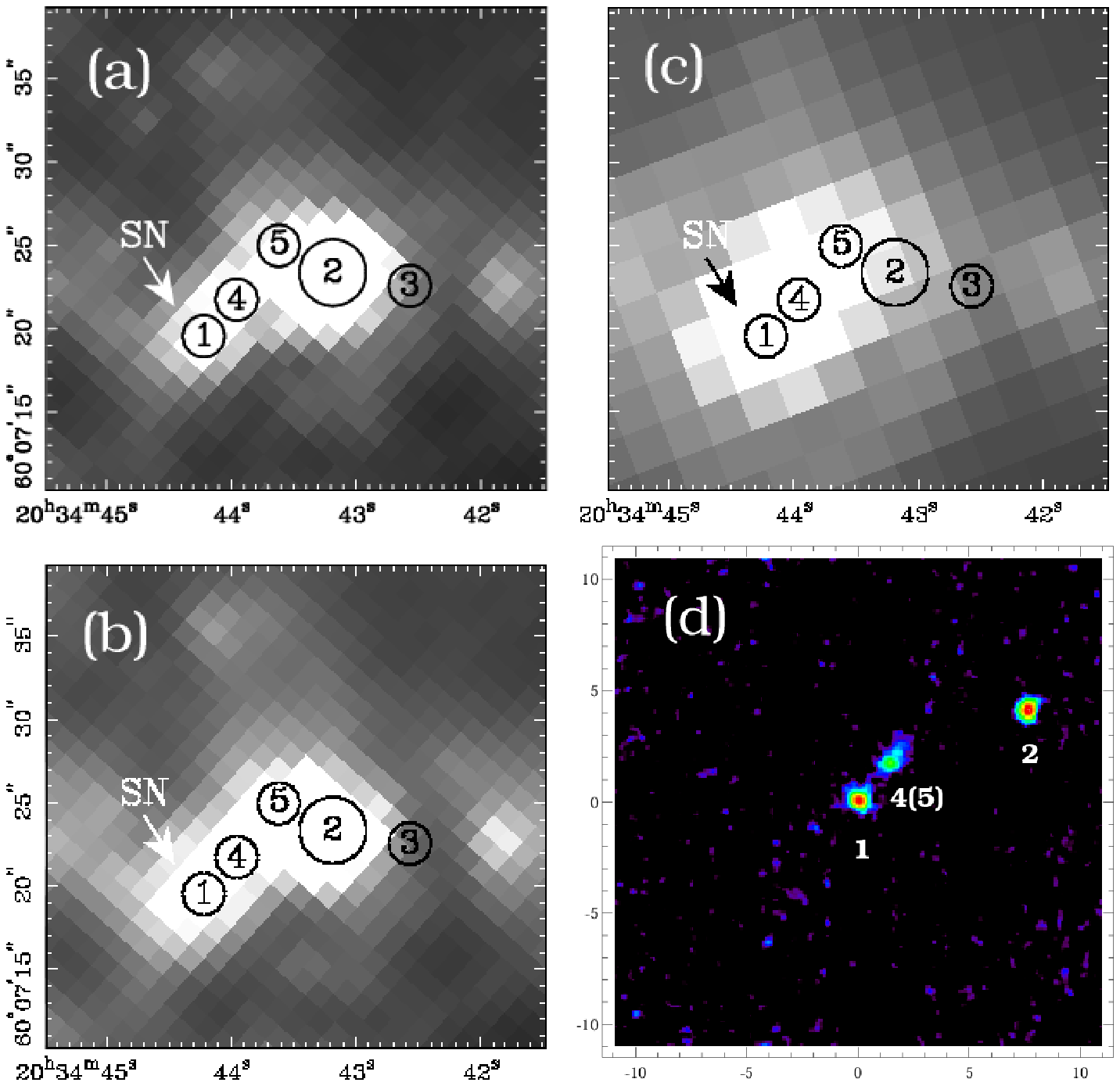}
\caption[]{
(a) and (b): {\em SST} SINGS IRAC 5.8-$\mu$m and 8.0-$\mu$m images
of a 30$''$ by 29$''$ region around SN~2002hh (pixel size = $1''.1$),
obtained on June 10th 2004. (c): {\em SST} SINGS MIPS 24-$\mu$m
image of the same region (pixel size = $2''.6$), obtained on July 9th
2004. (d): Gemini-N Michelle 11.2-$\mu$m image of a
$21''.8\times21''.8$ region centred on SN~2002hh ($0''.099$/pixel),
obtained on Sept 26th 2004. Offsets in arcseconds from the position of the
SN are marked on the axes. A 3-pixel ($0''.3$) FWHM Gaussian filter was
applied to the cleaned image.}
\end{figure}

\clearpage

\begin{figure}
\plotone{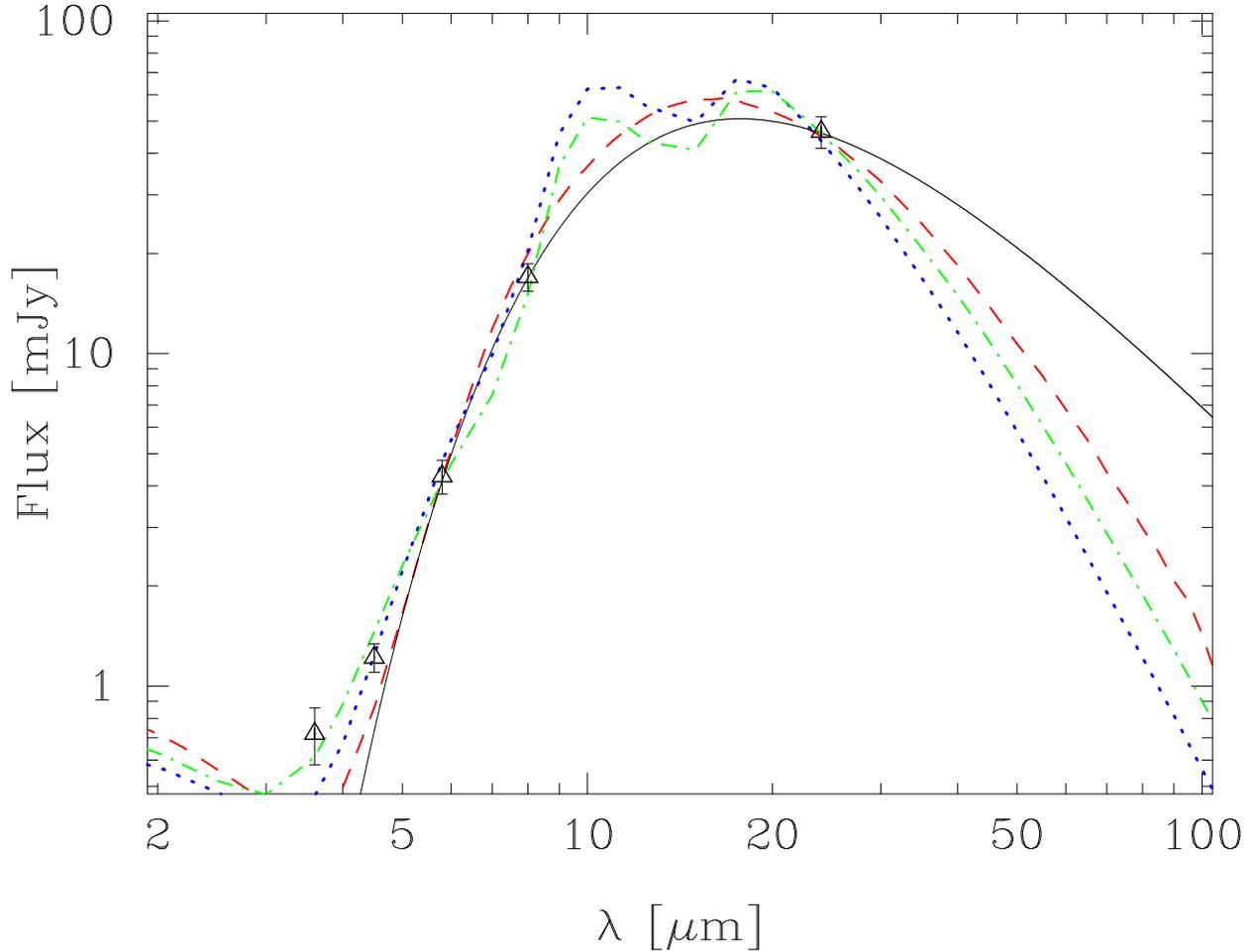}
\caption[]{The measured IRAC and MIPS fluxes from SN~2002hh in
June/July 2004 are shown as open triangles, with vertical bars indicating
the flux uncertainties. The solid black line is a 290-K blackbody
normalised to the 8.0-$\mu$m flux point. The dashed, dotted and
dash-dotted lines correspond
to radiative transfer dust models that are discussed in the text.
} 
\end{figure}

\clearpage

\begin{center}
\begin{table}
\caption{Fluxes from the {\em Spitzer Space Telescope}, in mJy}
\begin{tabular}{llllll}
\hline\hline	     
Instrument  &  IRAC  &  IRAC & IRAC & IRAC & MIPS \\
$\lambda$ [$\mu$m] & 3.6 & 4.5 & 5.8 & 8.0 & 24.0 \\
T(sec)$^a$ & 26.8 & 26.8 & 26.8 & 26.8 & 7.34 \\
\noalign{\vskip3pt} \noalign{\hrule} \noalign{\vskip3pt}
Star~1 Jun/Jul 04 & 0.72 $\pm$ 0.14 & 1.22 $\pm$ 0.12 & 4.28 $\pm$ 0.50 & 17.0 $\pm$ 1.6 & 46.5 $\pm$ 5.1 \\
Star~1 25th Nov 04     & 0.88 $\pm$ 0.17 & 1.07 $\pm$ 0.12 & 3.65 $\pm$ 
0.35 & 12.8 $\pm$ 1.3 &  ~~~~~~~-- \\
 & & & & & \\
Star~2 10th Jun 04 & 119.6 $\pm$ 18.4 & 68.9 $\pm$ 4.2  & 54.9 $\pm$ 2.1  
& 31.4 $\pm$ 1.6 & ~~~~~~~--  \\
Star~2 25th Nov 04     & 115.6 $\pm$ 15.4 & 58.8 $\pm$ 4.2  & 53.5 $\pm$ 
2.2 & 30.4 $\pm$ 1.2 & ~~~~~~~--  \\
 & & & & & \\
Star~3 10th Jun 04 & 1.12 $\pm$ 0.35 & 0.54 $\pm$ 0.10 & 0.71 $\pm$ 0.21 & 
1.23 $\pm$ 0.28 & ~~~~~~~-- \\
Star~3 25th Nov 04     & 0.78 $\pm$ 0.19 & 0.62 $\pm$ 0.21 & 0.58 $\pm$ 
0.16 & 1.78 $\pm$ 0.82 & ~~~~~~~-- \\
 & & & & & \\
Star~4 Jun/Jul 04 & 0.64 $\pm$ 0.17 & 0.57 $\pm$ 0.11 & 3.52 $\pm$ 0.46 & 9.3 $\pm$ 1.2  & 27.7 $\pm$ 4.5 \\
Star~4 25th Nov 04     & 0.64 $\pm$ 0.24 & 0.59 $\pm$ 0.13 & 3.27 $\pm$ 
0.36 & 10.9 $\pm$ 1.1  & ~~~~~~~--  \\
 & & & & & \\
Star~5 10th Jun 04 & ~~~~~~~-- & ~~~~~~~-- & 2.11 $\pm$ 0.36 & 8.6 $\pm$ 
1.2 & ~~~~~~~--  \\
Star~5 25th Nov 04     & ~~~~~~~-- & 0.18 $\pm$ 0.09 & 1.80 $\pm$ 0.32 & 
6.7 $\pm$ 1.1 & ~~~~~~~-- \\
\noalign{\vskip3pt} \noalign{\hrule}
\end{tabular}	        
\begin{description}
\item[$^a$] Effective exposure time per pixel
\end{description}
\end{table}
\end{center}

\end{document}